\documentclass[twocolumn]{aastex631}

\usepackage{appendix}

\newcommand{\oii}{\relax \ifmmode {\mbox O\,{\scshape ii}}\else O\,{\scshape ii}\fi}
\newcommand{\oiii}{\relax \ifmmode {\mbox O\,{\scshape iii}}\else O\,{\scshape iii}\fi}

\newcommand{\Halpha}{\hbox{H$\alpha$}}

\newcommand{\Hbeta}{\hbox{H$\beta$}}

\newcommand{\NII}{\hbox{[N\,{\sc ii}]}}
\newcommand{\SIII}{\hbox{[S\,{\sc iii}]$\lambda9069,9532$}}
\newcommand{\HeI}{\hbox{He\,{\sc i}$\lambda10830$}}
\newcommand{\OIII}{\hbox{[O\,{\sc iii}]}}

\begin{document}

\title{JEMS: A deep medium-band imaging survey in the {\it Hubble} Ultra-Deep Field with {\it JWST} NIRCam \& NIRISS}

\author[0000-0003-2919-7495]{Christina C. Williams}
\affiliation{NSF’s National Optical-Infrared Astronomy Research Laboratory, 950 N. Cherry Avenue, Tucson, AZ 85719, USA}
\affiliation{Steward Observatory, University of Arizona, 933 North Cherry Avenue, Tucson, AZ 85721, USA}

\author[0000-0002-8224-4505]{Sandro Tacchella}
\affiliation{Kavli Institute for Cosmology, University of Cambridge, Madingley Road, Cambridge, CB3 0HA, UK}
\affiliation{Cavendish Laboratory, University of Cambridge, 19 JJ Thomson Avenue, Cambridge, CB3 0HE, UK}

\author[0000-0003-0695-4414]{Michael V. Maseda}
\affiliation{Department of Astronomy, University of Wisconsin-Madison, 475 N. Charter St., Madison, WI 53706 USA}

\author[0000-0002-4271-0364]{Brant E. Robertson}
\affiliation{Department of Astronomy and Astrophysics, University of California, Santa Cruz, 1156 High Street, Santa Cruz, CA 95064, USA}

\author[0000-0002-9280-7594]{Benjamin D. Johnson}
\affiliation{Center for Astrophysics $\vert$ Harvard \& Smithsonian, 60 Garden Street, Cambridge, MA 02138, USA}

\author[0000-0002-4201-7367]{Chris J. Willott}
\affil{NRC Herzberg, 5071 West Saanich Rd, Victoria, BC V9E 2E7, Canada}

\author[0000-0002-2929-3121]{Daniel J. Eisenstein}
\affiliation{Center for Astrophysics $\vert$ Harvard \& Smithsonian, 60 Garden Street, Cambridge, MA 02138, USA}

\author[0000-0001-9262-9997]{Christopher N. A. Willmer}
\affiliation{Steward Observatory, University of Arizona, 933 North Cherry Avenue, Tucson, AZ 85721, USA}

\author[0000-0001-7673-2257]{Zhiyuan Ji}
\affiliation{Steward Observatory, University of Arizona, 933 North Cherry Avenue, Tucson, AZ 85721, USA}

\author[0000-0001-9262-9997]{Kevin N. Hainline}
\affiliation{Steward Observatory, University of Arizona, 933 North Cherry Avenue, Tucson, AZ 85721, USA}

\author[0000-0003-4337-6211]{Jakob M. Helton}
\affiliation{Steward Observatory, University of Arizona, 933 North Cherry Avenue, Tucson, AZ 85721, USA}

\author[0000-0002-8909-8782]{Stacey Alberts}
\affiliation{Steward Observatory, University of Arizona, 933 North Cherry Avenue, Tucson, AZ 85721, USA}

\author{Stefi Baum}
\affiliation{Dept of Physics \&
Astronomy University of Manitoba, 30A Sifton Rd, Winnipeg MB R3T 2N2 Canada.}

\author[0000-0003-0883-2226]{Rachana Bhatawdekar}
\affiliation{European Space Agency (ESA), European Space Astronomy Centre (ESAC), Camino Bajo del Castillo s/n, 28692 Villanueva de la Cañada, Madrid, Spain}

\author[0000-0003-4109-304X]{Kristan Boyett}
\affiliation{School of Physics, University of Melbourne, Parkville 3010, VIC, Australia}
\affiliation{ARC Centre of Excellence for All Sky Astrophysics in 3 Dimensions (ASTRO 3D), Australia}

\author{Andrew J. Bunker}
\affiliation{Department of Physics, University of Oxford, Denys Wilkinson Building, Keble Road, Oxford OX13RH, U.K.}

\author[0000-0002-6719-380X]{Stefano Carniani}
\affiliation{Scuola Normale Superiore, Piazza dei Cavalieri 7, I-56126 Pisa, Italy}

\author[0000-0003-3458-2275]{Stephane Charlot}
\affiliation{Sorbonne Universit\'e, CNRS, UMR 7095, Institut d'Astrophysique de Paris, 98 bis bd Arago, 75014 Paris, France}

\author[0000-0002-7636-0534]{Jacopo Chevallard}
\affiliation{Department of Physics, University of Oxford, Denys Wilkinson Building, Keble Road, Oxford OX13RH, U.K.}

\author[0000-0002-9551-0534]{Emma Curtis-Lake}
\affiliation{Centre for Astrophysics Research, Department of Physics, Astronomy and Mathematics, University of Hertfordshire, Hatfield AL10 9AB, UK}

\author[0000-0002-2380-9801]{Anna de Graaf}
\affiliation{Max-Planck-Institut f\"ur Astronomie, K\"onigstuhl 17, D-69117, Heidelberg, Germany}

\author[0000-0003-1344-9475]{Eiichi Egami}
\affiliation{Steward Observatory, University of Arizona, 933 North Cherry Avenue, Tucson, AZ 85721, USA}

\author[0000-0002-8871-3026]{Marijn Franx}
\affiliation{Leiden Observatory, Leiden University, P.O.Box 9513, NL-2300 AA Leiden, The Netherlands}

\author[0000-0002-5320-2568]{Nimisha Kumari}
\affiliation{AURA for the European Space Agency, Space Telescope Science Institute, 3700 San Martin Drive, Baltimore, MD 21218, USA}

\author[0000-0002-4985-3819]{Roberto Maiolino}
\affiliation{Kavli Institute for Cosmology, University of Cambridge, Madingley Road, Cambridge, CB3 0HA, UK; Cavendish Laboratory, University of Cambridge, 19 JJ Thomson Avenue, Cambridge, CB3 0HE, UK}

\author[0000-0002-7524-374X]{Erica J. Nelson}
\affiliation{Department for Astrophysical and Planetary Science, University of Colorado, Boulder, CO 80309, USA}

\author[0000-0002-7893-6170]{Marcia J. Rieke}
\affiliation{Steward Observatory, University of Arizona, 933 North Cherry Avenue, Tucson, AZ 85721, USA}

\author[0000-0001-9276-7062]{Lester Sandles}
\affiliation{Kavli Institute for Cosmology, University of Cambridge, Madingley Road, Cambridge, CB3 0HA, UK}
\affiliation{Cavendish Laboratory, University of Cambridge, 19 JJ Thomson Avenue, Cambridge, CB3 0HE, UK}

\author[0000-0003-4702-7561]{Irene Shivaei}
\affiliation{Steward Observatory, University of Arizona, 933 North Cherry Avenue, Tucson, AZ 85721, USA}

\author[0000-0003-4770-7516]{Charlotte Simmonds}
\affiliation{Kavli Institute for Cosmology, University of Cambridge, Madingley Road, Cambridge, CB3 0HA, UK}
\affiliation{Cavendish Laboratory, University of Cambridge, 19 JJ Thomson Avenue, Cambridge, CB3 0HE, UK}

\author[0000-0001-8034-7802]{Renske Smit}
\affiliation{Astrophysics Research Institute, Liverpool John Moores University, 146 Brownlow Hill, Liverpool L3 5RF, UK}

\author[0000-0002-1714-1905]{Katherine A. Suess}
\affiliation{Department of Astronomy and Astrophysics, University of California, Santa Cruz, 1156 High Street, Santa Cruz, CA 95064 USA}
\affiliation{Kavli Institute for Particle Astrophysics and Cosmology and Department of Physics, Stanford University, Stanford, CA 94305, USA}

\author[0000-0002-4622-6617]{Fengwu Sun}
\affiliation{Steward Observatory, University of Arizona, 933 North Cherry Avenue, Tucson, AZ 85721, USA}

\author[0000-0003-4891-0794]{Hannah \"Ubler}
\affiliation{Kavli Institute for Cosmology, University of Cambridge, Madingley Road, Cambridge, CB3 0HA, UK; Cavendish Laboratory, University of Cambridge, 19 JJ Thomson Avenue, Cambridge, CB3 0HE, UK}

\author[0000-0002-7595-121X]{Joris Witstok}
\affiliation{Kavli Institute for Cosmology, University of Cambridge, Madingley Road, Cambridge, CB3 0HA, UK}
\affiliation{Cavendish Laboratory, University of Cambridge, 19 JJ Thomson Avenue, Cambridge, CB3 0HE, UK}

\begin{abstract}
We present JEMS ({\it JWST} Extragalactic Medium-band Survey), the first public medium-band imaging survey carried out using {\it JWST}/NIRCam and NIRISS. These observations use $\sim2\mu$m and $\sim4\mu$m medium-band filters (NIRCam F182M, F210M, F430M, F460M, F480M; and NIRISS F430M \& F480M in parallel) over 15.6 square arcminutes in the Hubble Ultra Deep Field (UDF), thereby building on the deepest multi-wavelength public datasets available anywhere on the sky. We describe our science goals, survey design,  NIRCam and NIRISS image reduction methods, and describe our first data release of the science-ready mosaics. Our chosen filters create a {\it JWST} imaging survey in the UDF that enables novel analysis of a range of spectral features potentially across the redshift range of $0.3<z<20$, including Paschen-$\alpha$, H$\alpha$+\NII, and \OIII+H$\beta$ emission at high spatial resolution. 
We find that our {\it JWST} medium-band imaging efficiently identifies strong line emitters 
(medium-band colors $>1$ magnitude) across redshifts $1.5<z<9.3$, most prominently H$\alpha$+\NII\ and \OIII+H$\beta$. We present our first data release including science-ready mosaics of each medium-band image available to the community, adding to the legacy value of past and future surveys in the UDF. We also describe future data releases. This survey demonstrates the power of medium-band imaging with {\it JWST}, informing future extragalactic survey strategies using {\it JWST} observations.
\end{abstract}


\section{Introduction} \label{sec:intro}

Optical and infrared extra-galactic deep-field surveys have revealed that during the first few billion years after the Big Bang, galaxies rapidly evolved under very different physical conditions than galaxies today. At fixed mass, early galaxies are smaller, have lower metal content and contain stars that produce 
harder ionizing radiation fields, driving strong ionized gas emission lines in their interstellar medium  \citep[ISM; e.g.][]{Strom2017,  Stark2016, Katz2023}. The extreme physical conditions inside young galaxies at early times are now thought to be capable of reionizing the intergalactic medium \citep[IGM; e.g.][]{Robertson2013, Bunker2010, Atek2015, Maseda2020, Matthee2022}.  Meanwhile, recent studies using spectroscopy have demonstrated that massive quiescent galaxies begin to emerge after only 1-2 billion years \citep[][]{Glazebrook2017, Schreiber2018, Forrest2020, Valentino2020, Nanayakkara2022}. These are the mature relics of rapid, extreme growth towards the end of the Reionization Era \citep[e.g.][]{Marrone2018,Williams2019,Wang2019,Sun2021,Long2022,Casey2019, Manning2022}. Even after the epoch of reionization completes, the majority of galaxies experience their most vigorous growth phases during the era of Cosmic Noon ($1<z<3$), forming new stars at unprecedentedly high rates across the universe \citep[e.g.][]{MadauDickinson2014}.

A better understanding of the physical drivers of early galaxy evolution requires data that accurately capture spectral features,  
which provide more robust indicators of galaxy properties, overcoming the degeneracies between redshift, age and dust reddening in spectral energy distributions (SEDs).
The unknown contribution of emission lines to broad-band fluxes has hampered high-redshift studies, and still impacts analysis using only {\it JWST} broad band photometry. While spectroscopic followup surveys have made important strides in characterizing spectral features for bright galaxies, major uncertainties remain for the fainter galaxies. An additional barrier to a complete picture of galaxy assembly is the incomplete knowledge about sub-kpc spectroscopic signatures within distant galaxies. Obtaining these data (e.g. with integral field spectroscopy) is extremely expensive at high-redshift, and remains limited to small samples of bright galaxies. Spatially resolved data from slitless spectroscopy or imaging with small bandwidths provide an efficient path forward by increasing efficiency and provide larger unbiased samples. However, slitless spectroscopy still typically have brighter detection limits than possible using imaging \citep[e.g.][]{Brammer2012,Colbert2013,Skelton2014}. Thus, imaging that finely samples galaxy SEDs at high spatial resolution has the power to identify the drivers of structural evolution across populations using statistical samples, identifying where star formation occurs at the time of observation, and where the stars have formed in the past on galaxy population scales.

\begin{figure*}[t]
\includegraphics[width=1\textwidth]{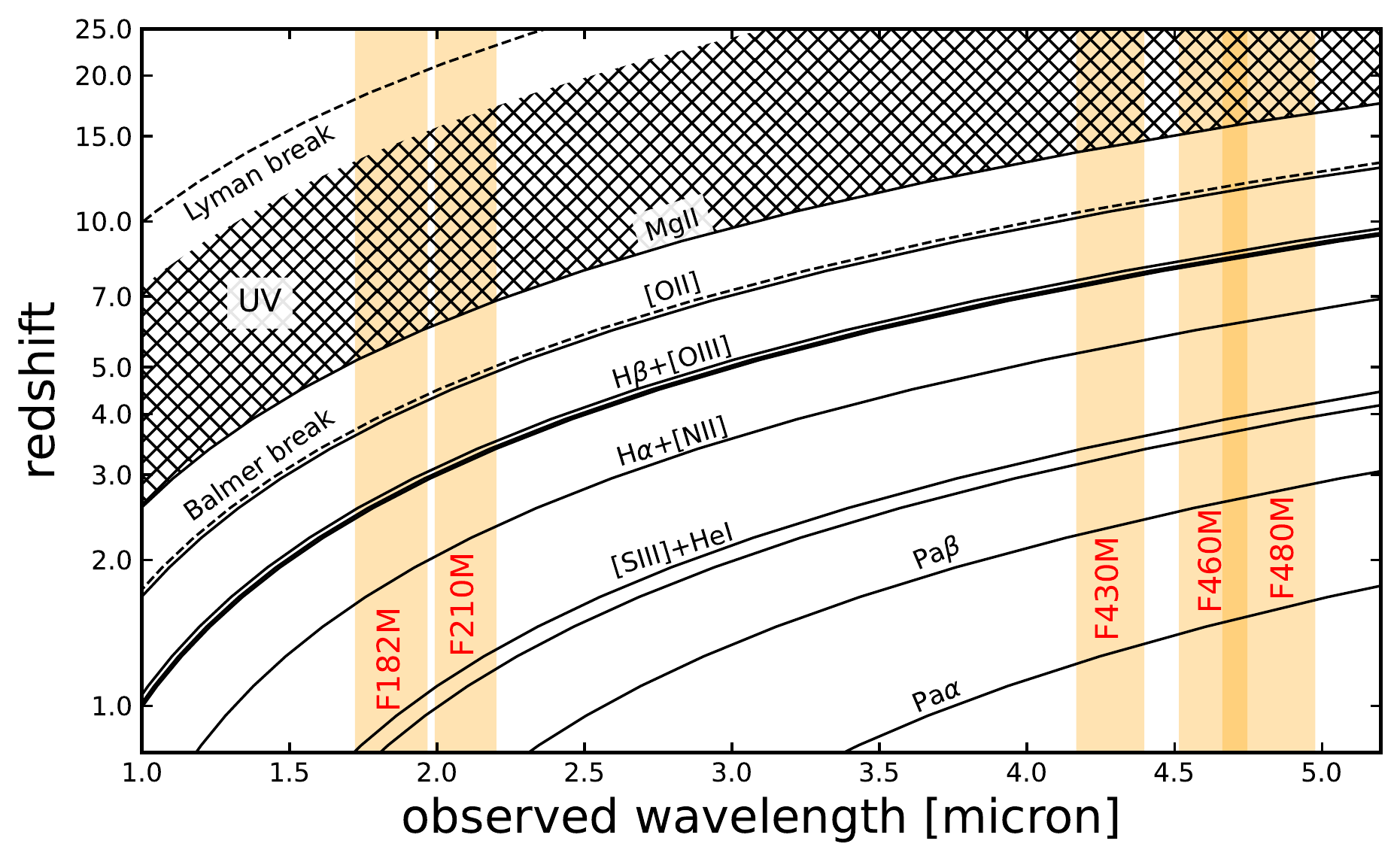}
\caption{Key spectral features as a function of redshift in the five medium-band filters (F182M, F210M, F430M, F460M, and F480M). Indicated are the Lyman break and Balmer break as dashed lines, the UV continuum ($1200-2800~\mathrm{\AA}$) as hatched band, and the various emission lines probed by our data.}\label{fig:lines}
\end{figure*}

An efficient path forward is to increase the sampling of SEDs by imaging with medium-band filters. Optical and near-infrared narrow- and medium-band surveys have demonstrated the power of increased spectral resolution to reconstruct galaxy properties out to wavelengths  $\lambda\lesssim2\mu$m, e.g. 
\citep{Wolf2004, Scoville2007, Moles2008, vanDokkum2009, Cardamone2010, Whitaker2011, PerezGonzalez2013,Straatman2016,Esdaile2021,Bonoli2021}. In particular, such surveys have decreased photometric redshift uncertainties to 1-2\% by spanning the redshifted Balmer / 4000\AA\ breaks for bright galaxies (to typical 5$\sigma$ detection limits of $<23-26$ ABmag). By doing so, these surveys have provided insight into the evolution of galaxy types by breaking the degeneracy between age and dust among red galaxy types out to $z\sim4$. Beyond this redshift, the bluest of strong rest-frame optical features, the Balmer / 4000\AA\ break, leaves the ground-based observational window ($\lambda\lesssim2\mu$m). Progress requires fainter detection limits and finer spectral sampling in key, unexplored wavelength ranges between 2-5$\mu$m. As an additional preview of the power of medium-bands at these wavelengths, the emission lines in early galaxies have been shown capable of boosting {\it Spitzer}/IRAC fluxes at $z>4$ despite very broad bandpasses \citep[see review in][and references therein]{Bradac2020}. These previous works motivate surveys using medium-bands at longer wavelengths to better break the degeneracies between continuum-break amplitude and line flux, and improve photometric redshifts.

Now that {\it JWST} has launched, it is revealing enormous discovery space across cosmic time \citep[see review by][]{Robertson2022}. A unique feature of {\it JWST} among space telescopes is the suite of medium-band filters spanning 1-5$\mu$m on board its near infrared camera \citep[NIRCam;][]{Rieke2005, Rieke2022}, and its near infrared imager and slitless spectrograph \citep[NIRISS;][]{Doyon2012, Willott2022}, enabling improved spectral sampling that can both measure and spatially resolve individual spectral features such as emission lines and continuum breaks. This capability not only enables serious improvements to redshift measurements, but the unprecedented sensitivity of {\it JWST} also enables much smaller uncertainties in inferring fundamental parameters of galaxies, 
to much fainter magnitudes than previously possible across all redshifts. Simulations of legacy deep field data have demonstrated the power of including one medium-band filter among the suite of broad band filters to recover intrinsic galaxy properties \citep[][]{Kemp2019, Kauffmann2020, RobertsBorsani2021, CurtisLake2021, Tacchella2022b}. As such, many extragalactic surveys planned in {\it JWST} Cycle 1 include 1-2 medium-bands to improve redshifts and SED measurements, including JADES (Eisenstein et al. in preparation), CEERS \citep{Bagley2022}, PRIMER \citep{Dunlop2021}, PANORAMIC \citep{WilliamsOesch2021}, UNCOVER \citep{Bezanson2022}, PEARLS \citep{Windhorst2022}, or even full suites of medium-band filters \citep[e.g. CANUCS][]{Willott2022}.

\begin{figure*}[!th]
   \includegraphics[scale=.46]{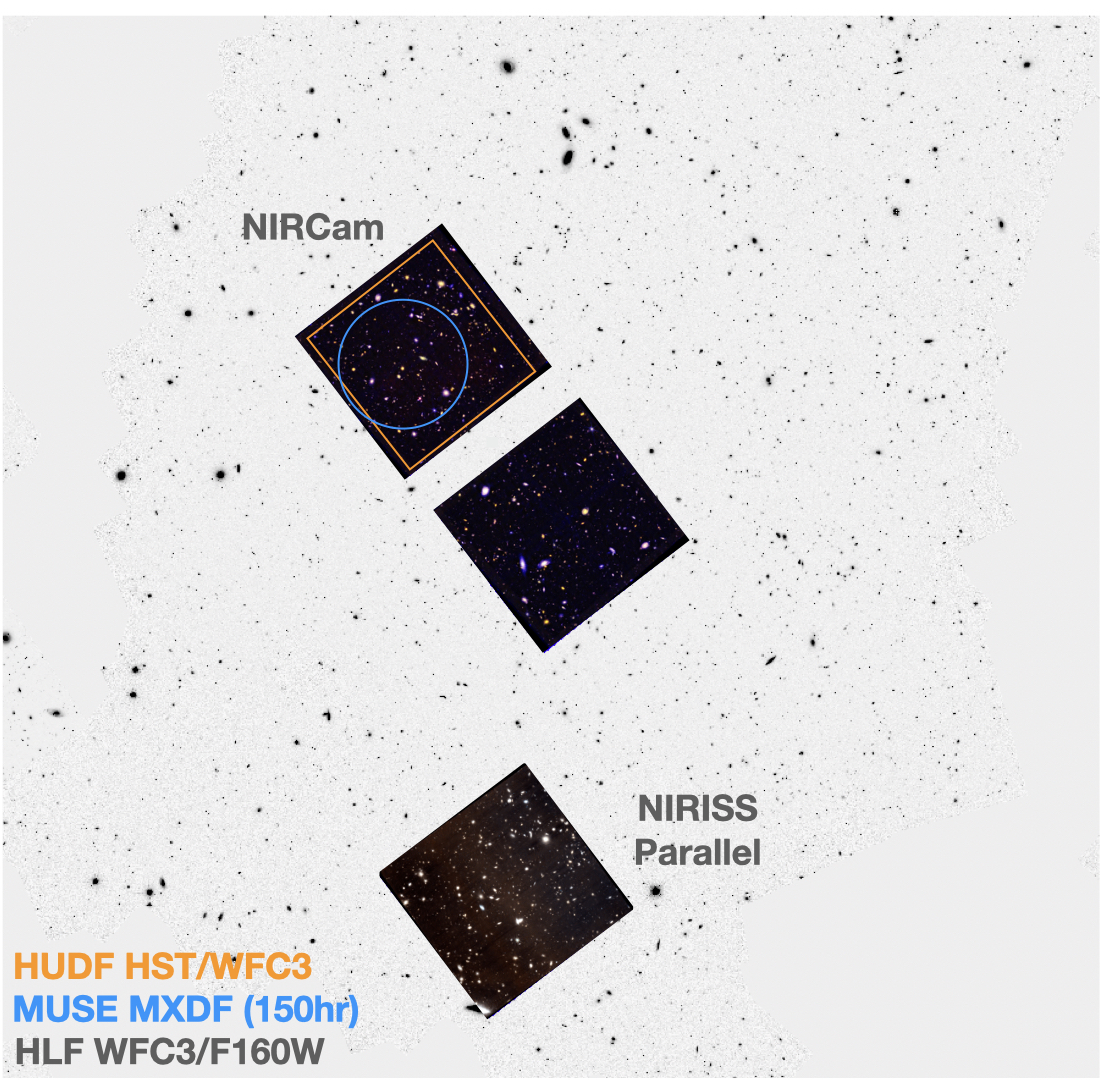}
  \caption{Layout of our NIRCam and NIRISS pointings in the GOODS-S field (HLF WFC3/F160W imaging, inverted grayscale, corrected to the GAIA astrometry). Colored regions indicate the footprint of various legacy data in the HUDF, covered by our NIRCam Module A:  HUDF WFC3/IR (orange), and MUSE MXDF (blue).  RGB mosaics of our JWST footprints are B:F430M G:F460M R:F480M for NIRCam, and B:F430M, G:F430M+F480M, R:F480M for NIRISS.}\label{fig:RGB}
\end{figure*}

\begin{figure*}[!th]
   \includegraphics[scale=.2]{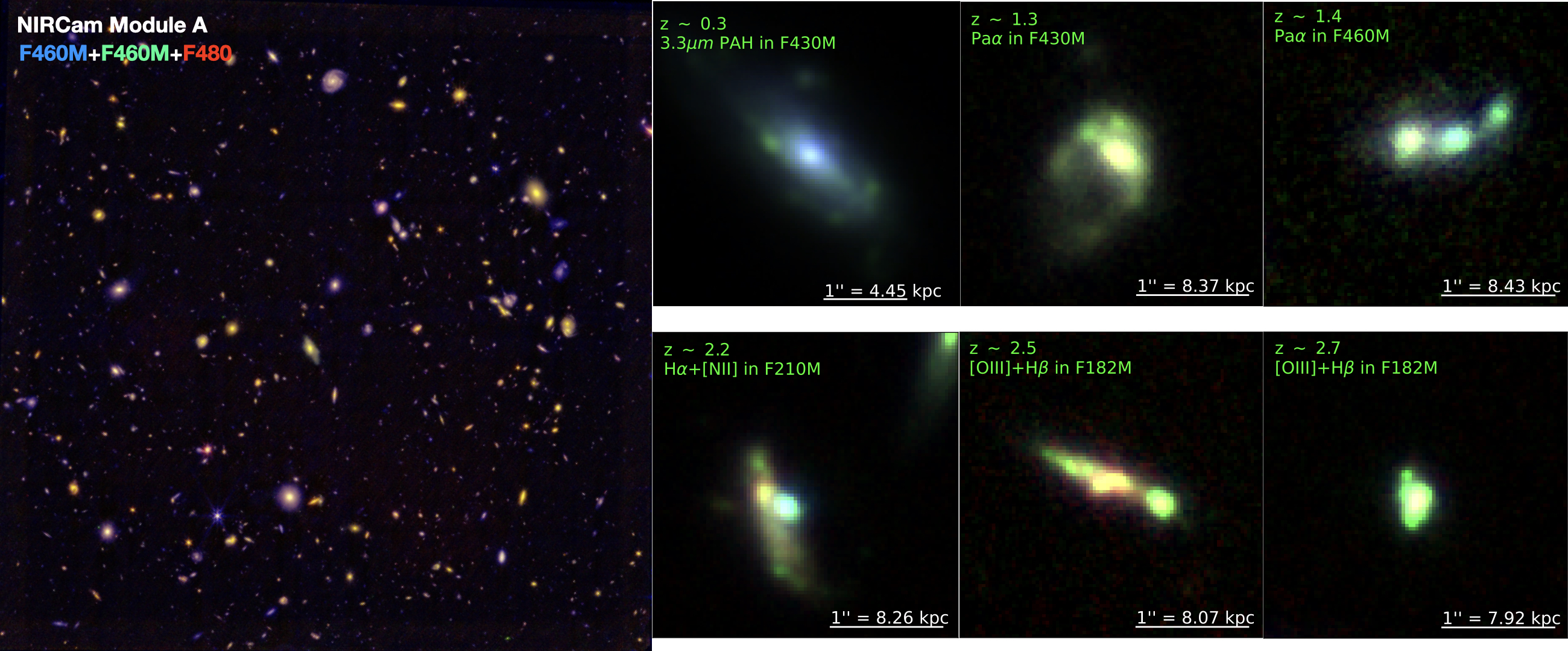}
     \includegraphics[scale=.2]{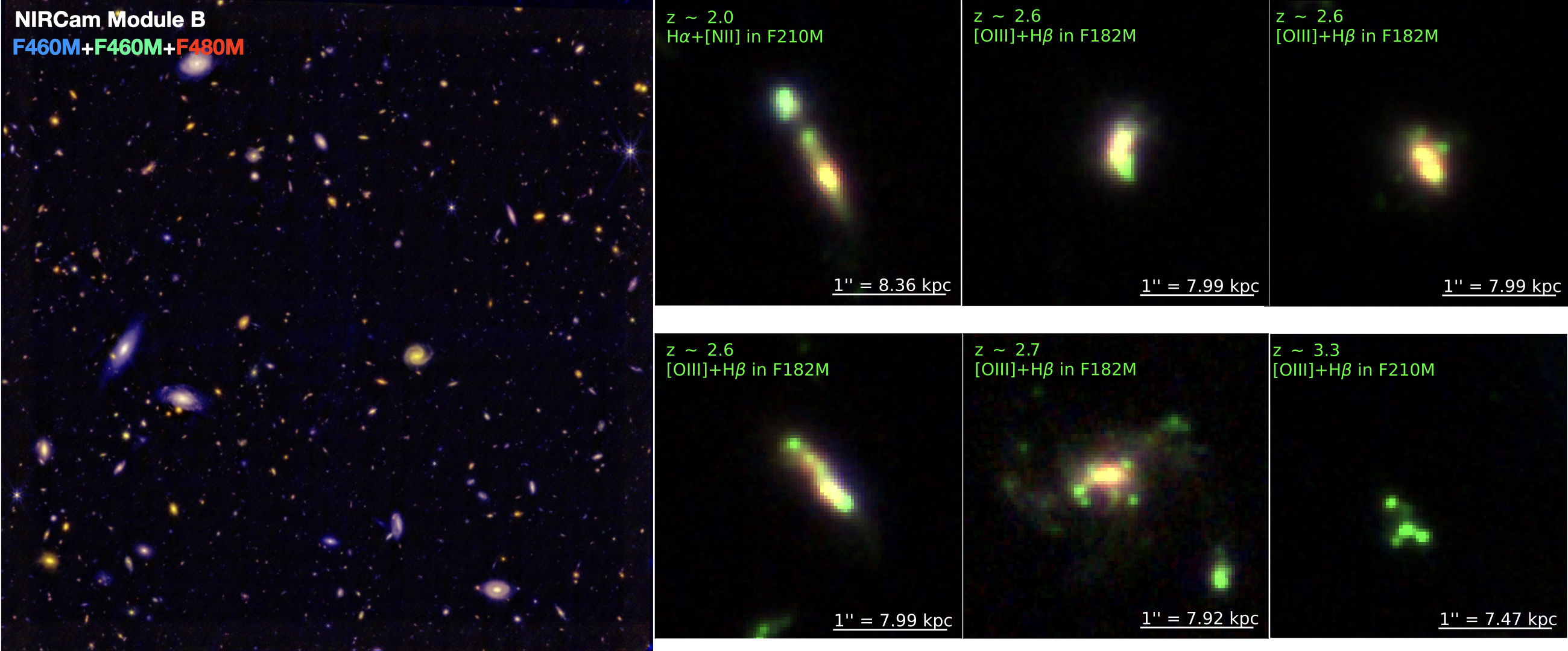}
       \includegraphics[scale=.2]{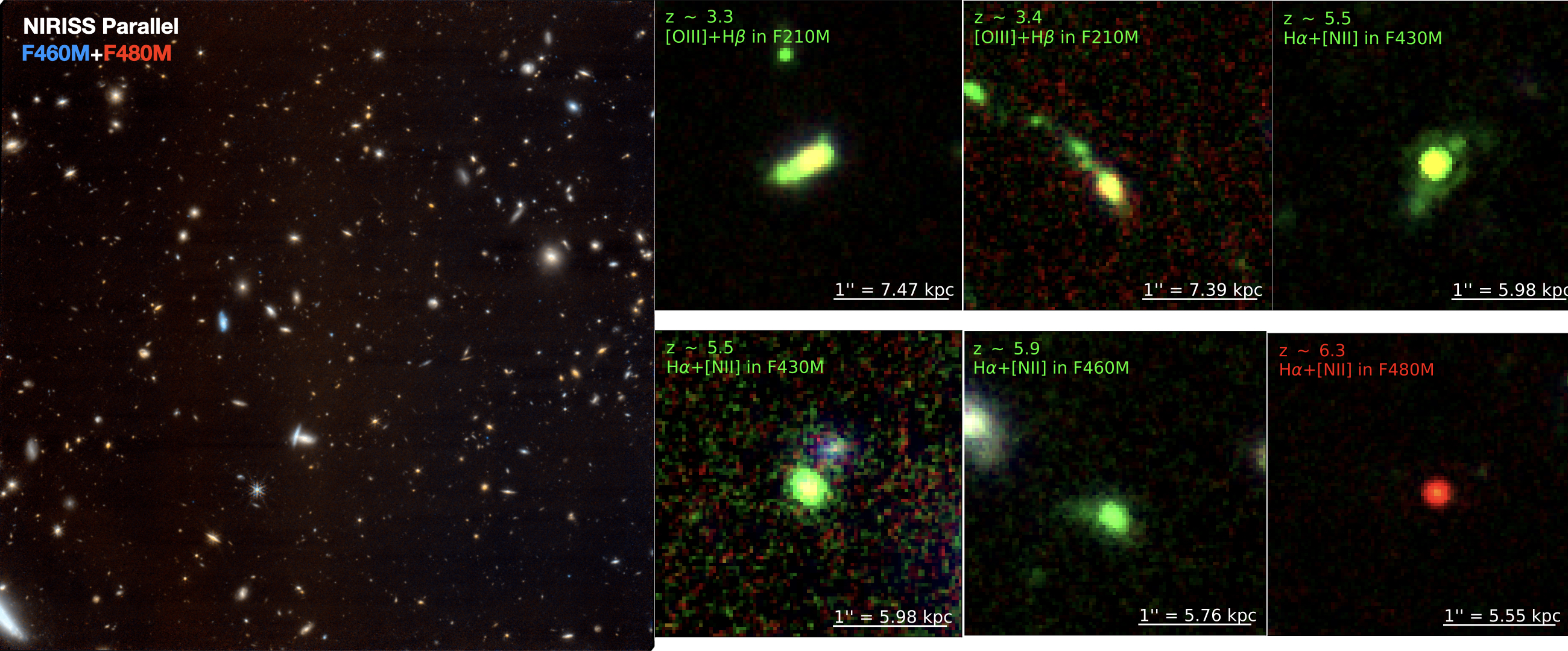}
  \caption{Left: RGB images for our three JWST pointings (as in Figure \ref{fig:RGB}. Right: image cutouts of spectroscopically confirmed galaxies at various redshifts whose colors trace emission line features (in order of increasing redshift). Blue: WFC3/160W; Red:F480M; Green:JWST filter where the corresponding emission line is. For F480M line emitters we use the F460M as green in the RGB.}\label{fig:RGB2}
\end{figure*}

\begin{deluxetable*}{llll}[!htbp]
\caption{Properties of our medium-band mosaics for NIRCam and NIRISS. Point source sensitivity is measured using 0.3" diameter apertures, aperture corrected to total magnitudes as described in Section \ref{sec:photometry}. }\label{tab:survey}
\tablehead{\colhead{Filter}  & \colhead{integration time} & \colhead{5$\sigma$ sensitivity}  & \colhead{Survey Area}} 
\startdata
NIRCam & seconds & ABmag &   10.1 sq arcmin   \\
\hline
F182M & 27830 & 29.3 &  \\
F210M & 27830 & 29.2   & \\
F430M & 13915 & 28.5  &\\
F460M & 13915 & 28.3  & \\
F480M & 27830 & 28.6  & \\
\hline
NIRISS  & & &  5.5 sq arcmin \\
\hline
F430M & 27057 &  28.4 &  \\
F480M & 27057 &  28.2 &  \\
\enddata
\end{deluxetable*}

This paper describes the {\it JWST} Extragalactic Medium-band Survey (JEMS), the first public imaging survey using more than two medium-band filters on board {\it JWST} at 2 to 5 $\mu$m. Our survey makes use of five medium-bands at key wavelength ranges (2 and 4$\mu$m) where increased sampling of the spectral energy distribution of high redshift galaxies can break important degeneracies in galaxy properties (see Figure \ref{fig:lines}). Thus, these data will improve the power of {\it JWST} data to reveal the physical processes of galaxy evolution \citep[e.g.][]{CurtisLake2021}, to faint limits below typical spectroscopic surveys even in the era of {\it JWST} \citep[see][]{Chevallard2019} and at high spatial resolution {\it for the first time}. In Section \ref{sec:obs} we outline our survey strategy and the JEMS data.  
In Section \ref{sec:data} we describe the data reduction procedure for each {\it JWST} instrument we use: NIRCam and NIRISS. In Section \ref{sec:image} we characterize our image characteristics and describe our procedure for photometric measurements. We conclude in Section \ref{sec:science} with the science drivers that motivated the design, and a demonstration of 
the science potential enabled by this data.

\section{Observations} \label{sec:obs}

\subsection{Survey Design}

The observations for our program (PID 1963, PIs C. Williams, S. Tacchella, M. Maseda) were conducted on Oct 12, 2022. Our observations include three separate footprints in the GOODS-S field \citep[][]{Giavalisco2004}. These include a single NIRCam pointing  in the Ultra Deep Field (UDF; \citealt{Beckwith2006}, NIRCam detector centered at RA=03:32:34, DEC=-27:48:08), composed of two individual footprints made by modules A and B. This pointing with an orient angle of 307.228 degrees aligns NIRCam Module A with the coverage of the deepest MUSE and {\it HST} pointings and covers the whole UDF \citep{Illingworth2016, Bacon2022}. The second NIRCam module (B) lies in the surrounding region covered by {\it HST} by both GOODS and CANDELS \citep{Grogin2011}, which also has some of the deepest MUSE and ALMA deep-field coverage on the sky to date. The entire NIRCam footprint is covered by ancillary near-infrared imaging from 0.9-4.4$\mu$m from the JADES survey (Eisenstein et al. in preparation). 
The observations consist of a single visit with three filter pairs: F210M-F430M (13915 sec), F210M-F460M (13915 sec), and F182M-F480M (27830 sec). We employ the DEEP8 readout pattern for all exposures/dithers (12 in total for F430M and F460M, 24 for F182M, F210M and F480M) with 6 groups per integration. 

A third footprint comes from performing a coordinated parallel with NIRISS imaging (using F430M and F480M filters), which increases the survey area at 4.3 and 4.8$\mu$m wavelengths by 50\%. The NIRISS footprint has a pointing center at 03:32:34, -27:54:01. We use the NIS readout pattern with 26 groups per integration, with total integration time of 28087 seconds for the F430M and F480M filters each. 
The NIRISS pointing sits inside CANDELS, and will have $\sim50$\% coverage by the JADES near-infrared imaging by the end of 2023. The location of our imaging footprints within the GOODS-S field is shown in Figure \ref{fig:RGB}. Our total on sky area covered by both NIRCam and NIRISS is $\sim$15.6 square arcminutes.

The dithering pattern is fixed to INTRAMODULEBOX with 4 primary dither positions and a 3-POINT-MEDIUM-WITH-NIRISS for the subpixel dither type. We choose the INTRAMODULEBOX pattern since it is more compact than INTRAMODULE or INTRAMODULEX, thereby yielding more area at full depth. The science exposure time is 15.47 hrs, and the total charged time is 20.42 hrs. Details of the observations using both instruments are listed in Table \ref{tab:survey}.

Our exposure times are motivated by probing the H$\alpha$ emission in $z = 5.4-6.6$ galaxies, both on integrated and spatially resolved scales. We aimed to obtain a point source line flux of $1-2 \times$ 10$^{-18}$ erg/s/cm$^2$, corresponding to a star formation rate (SFR) of $\sim$1 M$_{\odot}$/yr, which probes a major part of the galaxy population at this epoch (see Section \ref{sec:simulations}). In addition, we want to probe an \Halpha\ surface brightness of 4$\times10^{-18}$ erg/s/cm$^2$/arcsec$^2$, which enables a comparison of azimuthally-averaged \Halpha\ profiles to the typical Ly-$\alpha$ profiles out to the typical distance scales probed by MUSE \citep{Wisotzki2018}.

\begin{figure*}[!t]
    \includegraphics[width=1\textwidth,trim=90 0 50 0, clip]{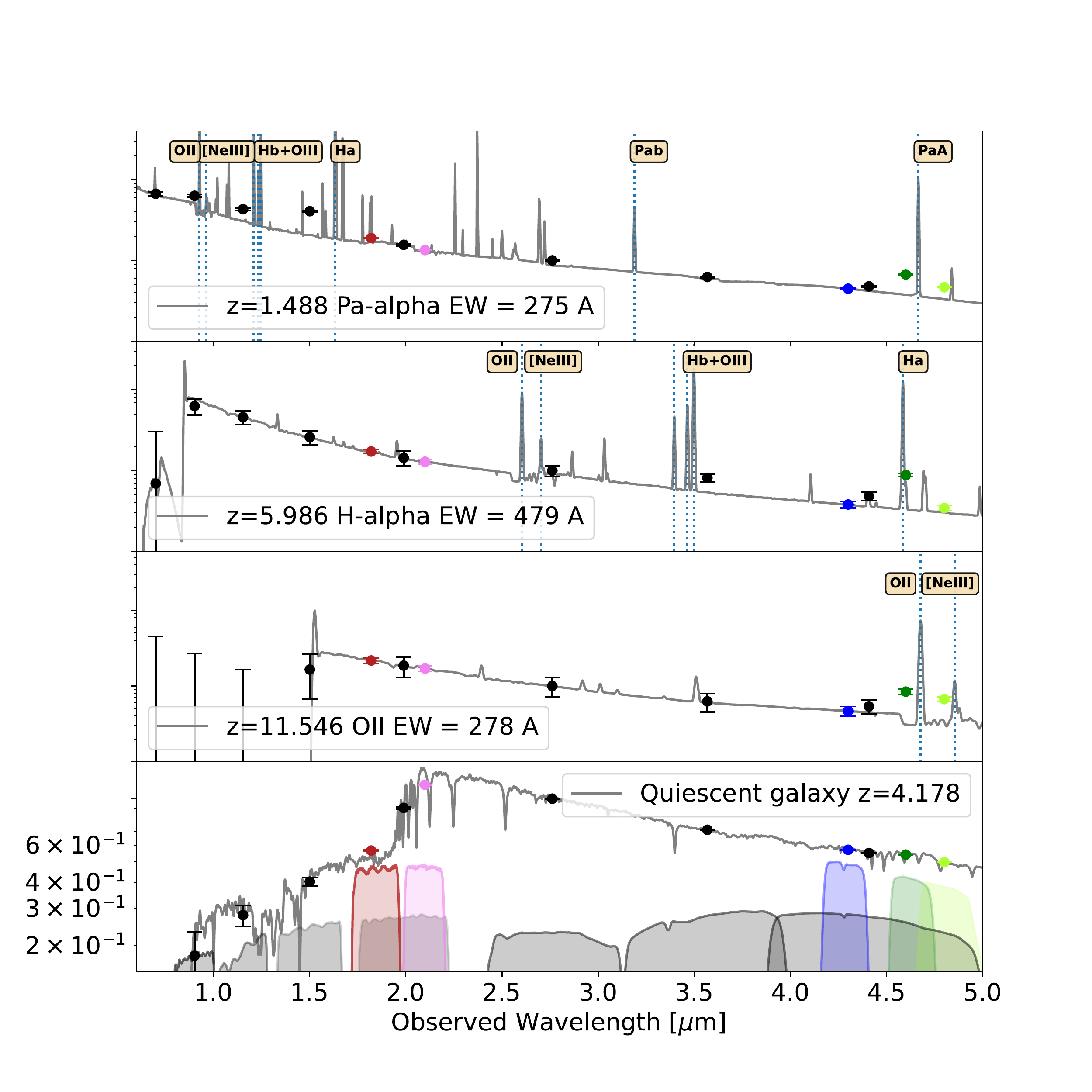}
  \caption{Example mock SEDs from JAGUAR \citep{Williams2018} demonstrating the power of our medium-band imaging (colored transmission curves and photometric points) to measure spectral features. SEDs are normalized at F277W (relative flux units). Top two panels include star forming galaxies with key redshifts and spectral features: Paschen-$\alpha$ at $z\sim1.4$, and \Halpha\ at $z\sim6$  (both seen with emission line in F460M). Third panel shows a $z\sim11.5$ galaxy with \oii\ flux boosting F460M. Bottom panel shows a quiescent galaxy at $z\sim4.1$. }\label{fig:SEDs}
\end{figure*}

Further, our medium-band survey with {\it JWST} was motivated as an important complement to ongoing spectroscopic surveys in the UDF, including with grism (e.g. FRESCO; \citealt{Oesch2021}; NGDEEP \citealt{Finkelstein2021}) and multi-object spectrograph \citep[e.g. JADES;][Eisenstein et al. in preparation;]{CurtisLake2022, Robertson2022c} and other spectroscopic programs \citep[e.g.][]{Kassin2021}. Medium-band imaging has more sensitive detection limits compared to grism spectroscopy per unit exposure time, because the grism reduces the throughput, covers a smaller effective area with both imaging and spectra per pointing, and for some galaxies with high velocity spread, resolves the line flux over more pixels.
A major technical challenge for grism observations involves overlapping source contamination along with blurring of the galaxy structure in the dispersion direction. Medium-band imaging is thus an incredibly powerful complementary dataset to help disentangle sources and reconstruct spatially resolved emission lines.  
Further, the medium band photometry can constrain the slit losses in NIRSpec emission line measurements, a major uncertainty in particular for the multi-shutter array, whose fixed grid of slitlets means galaxies are not always centered in their slits \citep{Ferruit2022}.

\subsection{Ancillary data }\label{sec:ancillary}

To characterize the sources detected in our survey later in this work (see Section \ref{sec:empirical}), we combine our imaging with existing ancillary data,
which is the deepest among all legacy fields in the sky. 
In particular, we make use of the Hubble Legacy Field  \citep[HLF;][and references therein]{Illingworth2016, Whitaker2019} imaging in GOODS-S from Hubble Space Telescope, which represents the deepest composite imaging including nine filters between 0.4-1.6$\mu$m wavelength (F435W, F606W, F775W, F814W, F850LP, F105W, F125W, F140W, F160W). We document our astrometric alignment procedure of these data with our JWST imaging in the Appendix. We also include photometry measured from the JADES NIRCam imaging obtained as of November 2022, which covers our NIRCam footprint (JADES data and processing methodology will be presented in detail in Eisenstein et al. in preparation). The JADES imaging in our footprint covers 9 filters across 0.9-4.4$\mu$m wavelengths (F090W, F115W, F150W, F200W, F277W, F335M, F356W, F410M, F444W). 

The HUDF is also home to extensive public and archival spectroscopy. We make use of a spectroscopic compilation of galaxies from the GOODS-S field as compiled in \citealt[][references therein]{Kodra2023} (their Table 6; N. Hathi, private communication).
We also include (and override the others in case of overlap) with the latest MUSE data release \citep{Bacon2022} and a preliminary list of spectroscopic identifications from FRESCO \citep{Oesch2021}, both of which target redshifts where our medium band imaging is most sensitive to line emitters ($z>3$). We only use redshifts which have been identified as secure or reliable and exclude uncertain or poor quality solutions.

\section{Data Reduction } \label{sec:data}

\subsection{NIRCam Image Reduction } \label{sec:datanrc}

Our NIRCam image processing follow the methodology developed for the JADES survey. Full details of the JADES imaging data reduction will be presented in Tacchella et al. (in preparation), including performance tests. We summarize here the main steps for completeness. We use version 1.8.1 of the Space Telescope Science Institute (STScI) {\it JWST} calibration pipeline. 
We use context map Calibration Reference Data System (CRDS) reference file {\tt jwst\_1007.pmap}. Importantly this context map includes updates from {\tt jwst\_0995.pmap}, which has the most recent absolute flux calibration for NIRCam detectors using our chosen medium-bands from observations collected by calibration programs PID 1536, 1537, and 1538 \citep[PI: K. Gordon][]{Boyer2022, Gordon2022}.

We run stage 1 of the pipeline with its default parameters. This stage performs detector-level corrections and produces count-rate images. We correct snowballs in the images caused by charge deposition following cosmic ray hits. We adopt the default values for stage 2 of the {\it JWST} pipeline, which performs the flat-fielding and applies the flux calibration. 

Following stage 2, we perform several custom corrections in order to account for several features in the NIRCam images \citep[e.g.][]{Rigby2022}. In particular, we remove the 1/f noise using a sigma-clipped median along the rows and then the columns on the source-masked images\footnote{\url{https://github.com/chriswillott/jwst}}. Following this, we subtract a smooth background image that we construct with the \texttt{photutils} \citep{photutils} Background2D class. For the short-wavelength channel images (in particular the NIRCam detectors A3, A4, B3 and B4 for the filters F182M and F210M) we simultaneously subtract scattered light artefacts \citep[i.e. ``wisps"][]{Rigby2022}. We have constructed wisp templates by stacking all images from our JADES (PID 1180) program and several other programs (PIDs 1063, 1345, 1837, 2738) after reducing them following the same approach as outlined above. The templates are rescaled to account for the variable brightness of the wisp features and then subtracted from the images. 

Before combining the individual exposures into a mosaic, we perform astrometric relative and absolute corrections with a custom version of {\it JWST} TweakReg. We group individual detector images by exposure number and then match sources to a reference catalog constructed from {\it HST} F160W mosaics in the GOODS-S field with astrometry tied to Gaia-EDR3 \citep[Brammer et al. in preparation;][]{GAIA2016, GAIA2018}. These matches are used to calculate a tangent plane shift and rotation for each exposure.  We then group the images by visit and band and apply the median shift and rotation for each group to all images in that group. At the time of writing, the flight versions of the distortion modeling for the medium-bands we used were not yet included in the CRDS. We override the default distortion model reference for the 2$\mu$m medium-band filters to use the distortion model for the nearest broad band filters (F200W for F182M and F210M, and F444W for F430M, F460M, F480M). We then run stage 3 of the {\it JWST} pipeline, combining all exposures of a given filter. We choose a pixel scale of 0.03 arcsec/pixel for both SW and LW channel images and choose the drizzle parameters \citep{Fruchter2002} of {\tt pixfrac}=1 and 0.7 for the 2$\mu$m and 4$\mu$m images, respectively.

\subsection{NIRISS Image Reduction } \label{sec:datanis}

The NIRISS images were processed with version 1.8.4 of the {\it JWST} calibration pipeline with CRDS context {\tt jwst\_1019.pmap}. Stage 1 of the pipeline included the optional snowball correction in the jump step. Two custom steps were run during stage 1: removal of random DC offsets along the detector columns using the NIRISS columnjump code\footnote{\url{https://github.com/chriswillott/jwst}} and flagging groups in pixels that are affected by persistence from the previous exposure. Stage 2 processing incorporated a 1/f noise correction on each flat-fielded rate file utilizing background subtraction and source masking to isolate the noise stripes from spatially-varying background or sources. A constant background model incorporating the NIRISS light saber scattered light feature (Doyon et al. submitted) was subtracted from each F430M exposure. The F480M data did not show a significant light saber feature. The twenty-four NIRISS exposures in each filter were then mosaicked with stage 3 of the pipeline utilizing an absolute astrometry reference catalog from JADES NIRCam imaging that overlaps almost half of the field (which are also astrometrically calibrated based on the same method described in Section 3.1).  Finally, a low level fitted 2D background model was subtracted from the mosaiced images. The NIRISS mosaic is drizzled onto a 30mas pixel scale.

\section{Image properties } \label{sec:image}

In this section, we characterize the properties of our final image mosaics including achieved depth, number of sources detected in each band, as well as in our stacked images of all bands combined. In order to accurately characterize our images, we first perform a simple analysis using source detection and then measure the photometry of those detected sources.

\subsection{Source detection} \label{sec:detection}

We perform source detection on our final mosaics. Detection images are constructed based on inverse variance weighting of {\it JWST}/NIRCam using SCI flux extension and ERR flux error extension (which includes sky, read and poisson noise added in quadrature) using astropy \citep{Astropy2022}. We create detection images for each of the five medium-band filters individually, and we also create a stacked detection image using the inverse variance weighted SCI and ERR image extensions. 
We identify detections as contiguous regions in the detection image with a minimum area of 10 pixels that contain signal to noise ratio SNR$>3$ using {\tt photutils} \citep{photutils}. We then apply a standard deblending algorithm 
to the detection image with parameters nlevels=32 and contrast=0.001 \citep{Bertin1996}. With these criteria, we detect 10686 objects in our stacked NIRCam imaging. Using detections in individual filters, we find  4876, 4528, 4505, 3391, 4876  sources (for F182M, F210M, F430M, F460M and F480M, respectively).
These include 106 sources that are detected in only 1 filter (5$\sigma$), representing candidates for strong emission line galaxies below the continuum detection limit. For the NIRISS footprint, we detect 2767 objects in F430M and 1976 in F480M.

\subsection{Photometry} \label{sec:photometry}

To characterize our survey performance, we use {\tt photutils} to create a simple photometric catalog by performing  forced photometry in fixed apertures of diameter 0.3" at the locations of our stacked detections, as well as for detections in the individual filters.
To measure photometric uncertainty in our forced apertures, we use 100,000 random apertures in sourceless regions identified using our source detection mask and measure the flux in the apertures, enabling us to include the additional noise source from pixel covariance in our uncertainty. We estimate aperture corrections for flux lost from the fixed apertures by interpolating encircled energy curves constructed using WebbPSF \citep{Perrin2014} for the nearest broad-band filter.

\subsection{Image results}

We measure the limiting $5\sigma$ depth of our images using the measured photometry of sources in 0.3" diameter apertures and the uncertainty measured in the random apertures described in the last section, which includes the impact of increased noise due to pixel covariance introduced by our mosaicking methodology. Thus our final mosaics are slightly shallower than the predictions from ETC (which excludes the uncertainty introduced by pixel covariance). The 5$\sigma$ magnitude limits (aperture corrected) in fixed apertures of 0.3" diameter in our 5 filters are presented in Table \ref{tab:survey}.

\begin{figure*}[!ht]
\includegraphics[width=1.\textwidth,trim=10 55 50 0, clip]{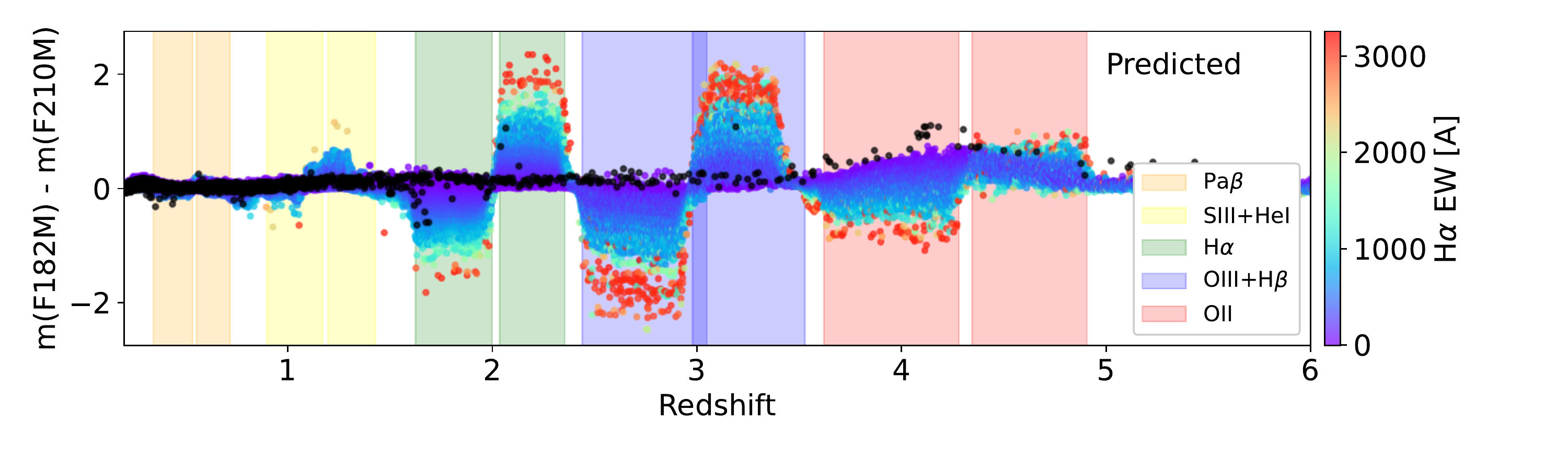}
\includegraphics[width=1.\textwidth,trim=10 20 50 15, clip]{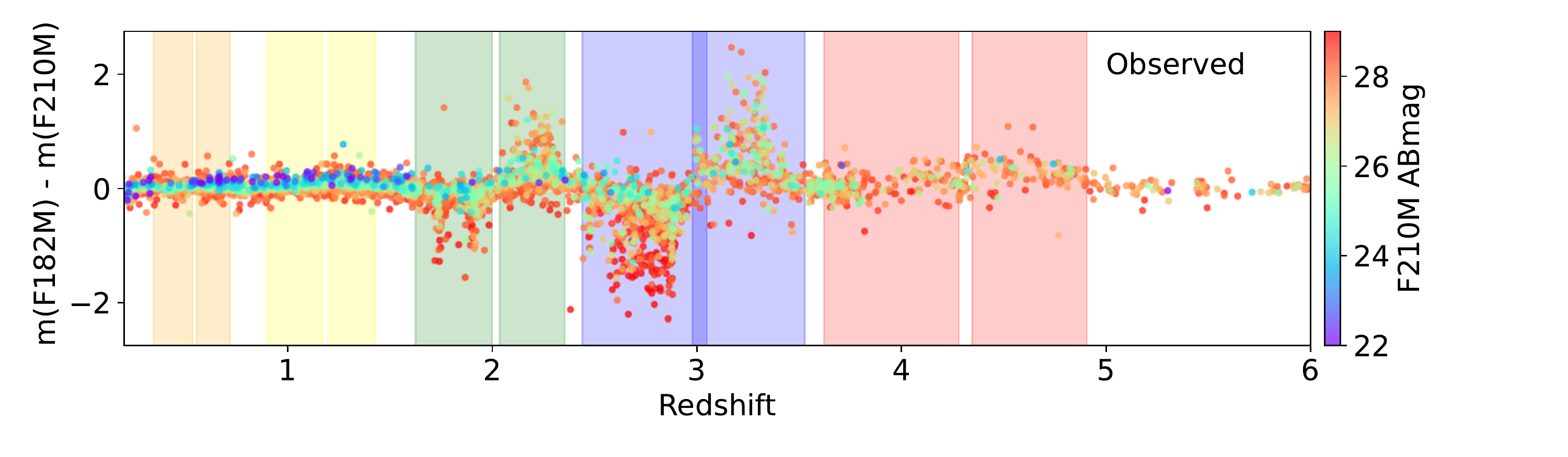}
\includegraphics[width=1.\textwidth,trim=10 55 50 0, clip]{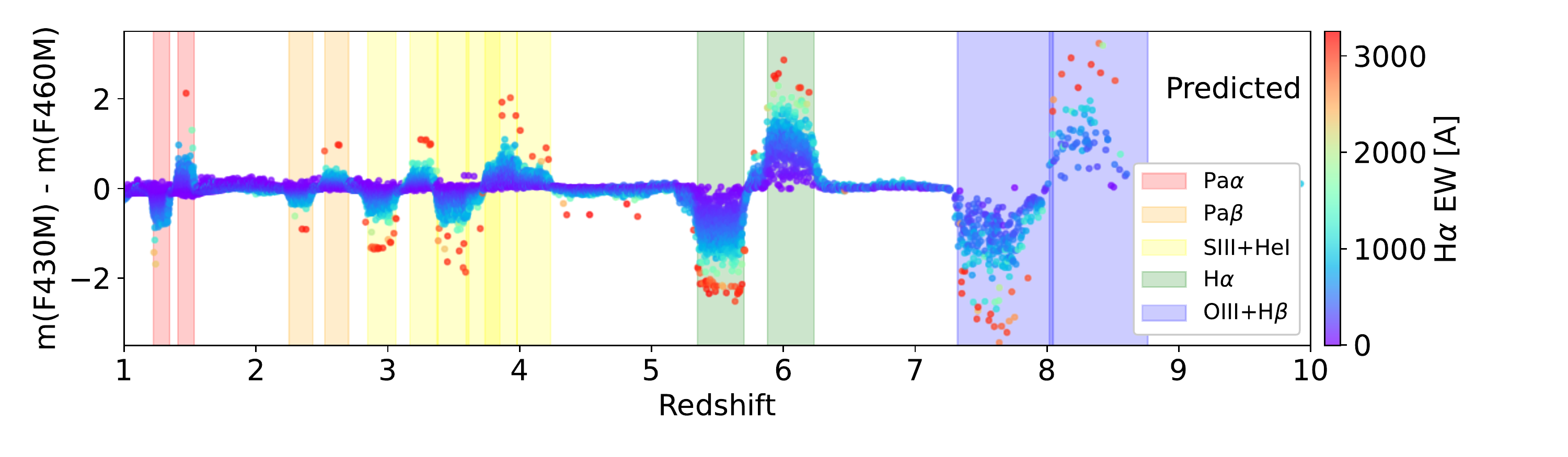}
\includegraphics[width=1.\textwidth,trim=10 20 50 15, clip]{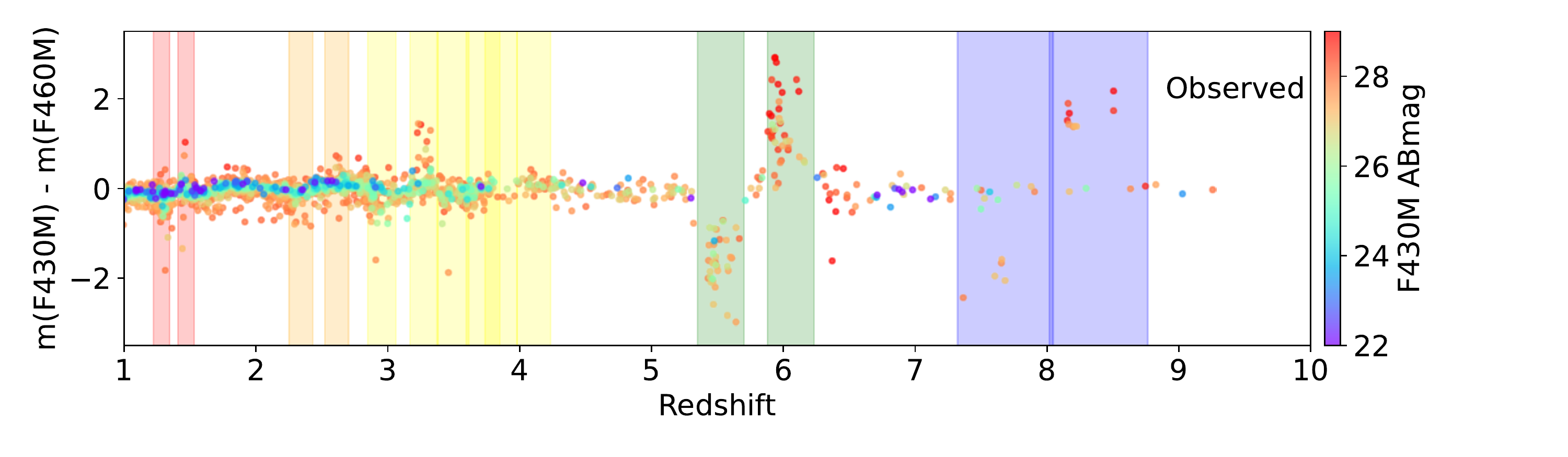}
\caption{SED shapes and strong emission lines are identifiable by extreme 2 and 4$\mu$m medium-band colors (e.g. F182M-F210M, top two panels, and F430M-F460M bottom two panels). In the first and third panels, points represent predicted distributions from mock star forming galaxies \citep{Williams2018} detected in either filter at $>7\sigma$ (flux limits in Table \ref{tab:counts}) color coded by their \Halpha\ rest-frame equivalent width (black points indicate quiescent galaxies). In the second and fourth panels, points are observed galaxies in our data, color-coded by their F430M apparent magnitude. In all panels, shaded bands indicate the emission feature boosting the flux at a given redshift. 2$\mu$m distributions show more sources than 4$\mu$m owing to the deeper detection limits. }\label{fig:MBcolors}
\end{figure*}

\section{Science  } \label{sec:science}

In this section, we present a view of the science potential of our medium-band imaging. We start by outlining the science drivers that motivated our survey design, in order of lookback time. We then present empirical findings for the galaxies for which we can measure emission line properties across redshifts, and outline our data's sensitivity to emission line properties  based on mock data.

\subsection{Science Objectives}

The new parameter space of spectral features observed at $\sim2$ and $\sim4$ $\mu$m wavelengths enables a wealth of new science across cosmic time.  Figure \ref{fig:lines} shows the observed wavelength of the major spectral features in galaxies for different redshifts, and where they cross into the medium-bands in our survey. New spectroscopic tracers are probed at nearly all redshifts from $0.3<z<20$. 
We structure the survey to allow exploration of the following science drivers.

\subsubsection{Star formation activity in galaxies during and after Cosmic Noon $(z<2.8)$}

One of the fundamental properties of galaxies is their star-formation rate (SFR). Despite significant evolution in galaxy properties over cosmic time, SFR measurements are still based on locally-calibrated diagnostics using nearby galaxies \citep{KennicuttEvans2012}. These efforts have resulted in the ``industry standard'' SFR indicator based on UV+IR, which is efficient for extra-galactic surveys out to $z\sim2$. 
However, traditional ``gold-standard" tracers that are least sensitive to dust attenuation have been inaccessible outside of the local universe until {\it JWST}. Measuring the instantaneous SFR from infrared Hydrogen recombination lines (e.g., Pa-$\alpha$) 
in both integrated and spatially resolved manner, is a major step forward in our understanding of galaxy assembly.

Our $4\mu$m imaging probes 3.3$\mu$m PAH emission at $z\lesssim1$ (one of the least-explored PAH bands beyond the local universe), Paschen lines Pa$\alpha\lambda18750$ and Pa$\beta\lambda12820$  at $z\sim1.2-2.8$, enabling less attenuation-affected SFR measurements for lines that have historically been only accessible at $z<0.3$ \citep[e.g.][]{Pasha2020, GimenezArteaga2022}. 
These less attenuation-affected SFR measurements have until now been only possible for relatively extreme sources at low redshift \citep{Calabro2018, Cleri2022} or rare highly magnified cases that are not necessarily representative of the full galaxy populations \citep[][]{Papovich2009, Rujopakarn2012}. While these lines are fainter than the Balmer series (even with no dust, Pa$\alpha/\mathrm{H}\alpha=0.33$ and Pa$\beta/\mathrm{H}\alpha=0.16$ for case B recombination), with the high sensitivity of {\it JWST}/NIRCam and NIRISS these measurements are in fact possible with medium-band photometry (see results presented in Section \ref{sec:empirical}).  Figure \ref{fig:SEDs} demonstrates an F460M flux excess due to Pa$\alpha$ emission in the SED of a mock star forming galaxy at $z\sim1.5$ from the JADES Extragalactic Ultra-deep Artificial Realization (JAGUAR) mock catalog \citep{Williams2018}, whose SEDs are created using BEAGLE \citep{ChevallardCharlot2016}. In principle, for galaxies that exhibit strong lines that boost colors with high enough signal to noise,  spatially resolved analyses can provide insight into how galaxies assemble their stellar mass (see Figure \ref{fig:RGB2}; Alberts et al. in preparation), building on previous studies with expensive IFU observations at the VLT \citep[e.g.][]{Tacchella2015, ForsterSchreiber2018} or {\it HST} grism \citep[e.g.][]{Nelson2016, Matharu2021}.

\subsubsection{Pre-Cosmic Noon $(2.5\lesssim z<5)$}

Prior to {\it JWST}, the ground-based observational limit at $\lambda\lesssim2.5\mu$m, and the low spatial and photometric resolution of {\it Spitzer}/IRAC, imposed technical challenges to understanding galaxy evolution above $z>3$. In particular, as the rest-frame optical diagnostics are redshifted beyond the high-resolution {\it HST} and deep ground-based K-band into broad IRAC bands, degeneracies are created between rest-frame optical emission line flux and continuum-break amplitude. These are the primary spectroscopic signatures that are leveraged for modeling the stellar populations in distant galaxies.

Our 2$\mu$m imaging traces the Balmer / 4000\AA\ break at $3.3<z<4.5$, which is now thought to be an important era when massive galaxies reach their maturity. The increased spectral sampling enables robust differentiation between SEDs that are red due to old age or dust content (see bottom panel of Figure \ref{fig:SEDs}). Ground-based limits made this only achievable for the most massive red galaxies (e.g. \citealt{Esdaile2021}), and {\it JWST} opens the window to explore quiescence among lower mass galaxies \citep[e.g.][]{Santini2022,Marchesini2023}. We thus expect that medium-band imaging at 2$\mu$m will be crucial to identifying the epoch of the emergence of massive quiescent galaxies and reconstructing their star-formation histories. The unprecedented spatial resolution will enable spatially-resolved color gradients of “dead” galaxies that trace age and metallicity gradients, measurements previously relegated to distant objects sheared by lensing \citep{Akhshik2020,Akhshik2022}. These empirical measurements provide powerful tests of the process that drives their rapid formation when compared to cosmological simulations \citep[e.g.][]{Wellons2015, Tacchella2016, Nelson2021}.

Among the most prominent features at this redshift range as seen in our 2$\mu$m data are the \OIII+H$\beta$ emission lines visible at $2.4 < z < 3.5$ (see e.g. top panel of Figure \ref{fig:MBcolors}). 
This new redshift coverage of \OIII+\Hbeta\ now enables continuous characterization of the \OIII+\Hbeta\ evolution between the reionization era 
and the analogs at $z\sim2-3$ that are typically used to infer broader conclusions about reionization physics \citep[e.g.][]{Nakajima2016, Fletcher2019,Barrow2020,Tang2021, Boyett2022}, thus tracking the abundance of strong line emitters and their equivalent width (EW) distribution among galaxy populations. Additionally, combined with the high resolution of NIRCam at 2$\mu$m, a new window is opened into the spatial distribution of star formation and ionized gas within galaxies during a critical assembly period (Ji et al. in preparation). 

We also note the potential discovery space of new diagnostics in this redshift range. As shown in Figure \ref{fig:MBcolors} (3rd panel), the 4$\mu$m medium-bands are expected to show strong color fluctuations with redshift due to the entrance and exit of \SIII+\HeI\  between $2.8 < z < 4.2$ based on the JAGUAR mock catalog \citep[][]{Williams2018}. These are not standard strong line diagnostics but they are strong enough to also be visible at Cosmic Noon in the 3D-{\it HST} grism survey \citep[][]{Momcheva2016}. While these features will undoubtedly improve photometric redshifts and the modeling of stellar populations in this redshift range, new science is possible through measuring their line fluxes via photometric excesses \citep[e.g.][]{Mingozzi2020}.

\subsubsection{End of Reionization  $(5.4<z<6.6)$}

Our data can ascertain if the flux excesses observed with {\it Spitzer} are due to Balmer breaks or due to contamination from strong rest-frame optical emission lines  \citep[e.g. H$\alpha$ EW$>$500\AA ;][]{Eyles2005,Eyles2007,Shim2011, Stark2013, deBarros2014,Smit2015,  Smit2016, Rasappu2016, Hatsukade2018, Lam2019, Faisst2019, Stefanon2022, Endsley2022,Sun2022}. Strong lines are reflective of bursty galaxy growth, compact stellar distributions and lower metallicity stars with harder ionizing radiation fields. 
Their properties are critical to understanding the process of Reionization  since star-forming galaxies are favored to drive it, though it is still debated which galaxies (bright versus faint) dominate \citep{Bunker2004,Robertson2013, Bouwens2015, Finkelstein2019, Naidu2020}. We still lack the necessary accounting of the production of ionizing photons from reionization era galaxies, and how they escape through the ISM and CGM. Importantly, Hydrogen-ionizing radiation will never be directly measured in the epoch of reionization (with mean redshift z$\sim7-8$, ending by z$\sim5-6$; e.g. \citealt{Planck2020, Keating2020}). This is due to the opacity of the intervening neutral IGM \citep[e.g.][]{Inoue2014, McGreer2015}, preventing an accurate measurement of the intrinsic ionizing photon production. The amount of dust attenuation and its geometry also remains significantly uncertain during these early phases of galaxy growth \citep[e.g.][]{Bowler2018,Bowler2022}.

Our three 4$\mu$m filters cover H$\alpha$+\NII\ at $5.4<z<6.6$ (the tail end of reionization epoch), simultaneously with improved sampling of the rest-frame UV continuum shape with the 2$\mu$m filters. The imaging thus provides critical constraints on the SFRs, stellar masses, intrinsic ionizing photon production and dust geometry within galaxies at high spatial resolution. Rest-frame optical emission line maps will create a wealth of diagnostics to enable a detailed picture of reionization era galaxies: the properties of the massive stars that ionize gas, where ionizing radiation initiates in the galaxy, and in combination with ancillary data, how Ly$\alpha$ radiation propagates to the CGM and IGM \citep[e.g.][Simmonds et al. in preparation, Maseda et al. in preparation]{Ning2022}. 

\subsubsection{Epoch of Reionization $(7.3<z<9.3)$}

{\it Spitzer}/IRAC photometric excesses have been observed, consistent with contamination from strong $[\mathrm{OIII}]$+H$\beta$ emission from redshifts $z\sim4-8$ \citep[e.g.][]{Labbe2013, DeBarros2019,Smit2014,Smit2015, Faisst2016,Endsley2021}
Above $z>7.3$ our 4$\mu$m imaging yields a direct measurement of $[\mathrm{OIII}]$+H$\beta$ at $z=7.6-9.3$, and the combination of the three filters enables a deconstruction of any flux from the Balmer break, probing earlier eras of star formation \citep[e.g.][]{Laporte2022}. 
This extends the accounting of ionizing photons and stellar populations deeper into the epoch of reionization, as strong $[\mathrm{OIII}]$+H$\beta$ lines are signposts for ionizing sources \citep[e.g.][]{Endsley2021}, while additionally providing inferences on the mass and redshift dependence of ionizing photon production as well as robust measurements of the mass function (otherwise contaminated by lines). The presence of emission lines probed by our medium-band photometry additionally can provide better redshift confirmation of lyman-break dropout samples identified with HST \citep[e.g.][]{Bunker2010, Wilkins2011,Bouwens2015,Finkelstein2015,Lorenzoni2013}.

\subsubsection{Cosmic Dawn $(9.3<z<20)$}

Limited samples of candidate galaxies with redshifts during Cosmic Dawn are known, with {\it JWST} now identifying photometric candidates beyond the redshift limit of {\it HST}+{\it Spitzer} \citep[$z\gtrsim11$;][]{Naidu2022,Atek2023,Finkelstein2022,Donnan2023,Harikane2022,Adams2023,Bradley2022}. Once $[\mathrm{OIII}]$+H$\beta$ redshifts outside of our 4$\mu$m medium-band window, our data provide  
access to novel spectral diagnostics for photometric data at these distant redshifts. These include the [OII]$\lambda\lambda$3727 line to $z\sim10.2-12.4$, whose typical line strengths are currently unknown at such redshifts (e.g. third panel of Figure \ref{fig:SEDs}). Further, our data enable the potential for detecting breaks at the Balmer limit \citep[rest-frame wavelength $\lambda3650$, probed by our data in the redshift range $z\sim 10.4 - 12.6$; see e.g.][]{ CurtisLake2022, Robertson2022c, Bouwens2022, Donnan2022b}, thus yielding the potential to reconstruct even yet earlier epochs of star formation \citep[if it occurs, see e.g. ][]{Labbe2022, Whitler2022, Tacchella2022d}.
Between the sets of 2 and 4$\mu$m filters, UV continuum colors are measurable to $z\sim9-23$. Further, we note the potential for refined dropout selection with narrower redshift selection function at $z\gtrsim14$ with the $2\mu$m filters.

\subsection{Empirical emission line constraints using medium-band imaging } \label{sec:empirical}

In this section we characterize the galaxy population for which our deep imaging  can provide new physical insight through refined SED sampling at unprecedented wavelength ranges (2-4$\mu$m). We focus specifically on our NIRCam photometry, which has the best ancillary data for assessing performance, given its location in the HUDF with the deepest {\it HST} legacy imaging, existing imaging from the JADES survey, and the largest sample of spectroscopy at the relevant redshifts (Section \ref{sec:ancillary}). 

To demonstrate the impact of spectral features (in particular emission lines) on the medium-band colors, we plot the 2$\mu$m colors (F182M-F210M) and one set of 4$\mu$m colors (F430M-F460M) as a function of redshift in Figure \ref{fig:MBcolors}. In the first and third panels, we show simulated color evolution with redshift using JAGUAR \citep{Williams2018}. This figure demonstrates that colors can reach more than 2 magnitudes difference depending on the intrinsic emission line properties of the galaxies (e.g. in the figure we color the points by their intrinsic H$\alpha$ equivalent width). This is by design much stronger than color evolution using {\it JWST} broad band filters \citep[colors of order $\sim$0.5 magnitudes as predicted by cosmological simulations, e.g.][]{Wilkins2022}. We also highlight the redshift ranges where major emission line features enter and exit the filter bandpasses as colored bands. The number of galaxies at higher redshifts starts to decrease (e.g. in particular at $z>7$, as can be seen in the 4$\mu$m colors in panel 3). Therefore, in order to highlight the features of the distributions at high-redshift, we maximize the number of galaxies in the distribution by plotting all JAGUAR sources down to stellar mass $>$ 10$^6$ M$_{\odot}$, and over 10 JAGUAR realizations ($\sim$1200 square arcmin). While this mock area is significantly larger than our imaging survey, it is useful as a demonstration of how the medium-band colors scale with emission line equivalent width. Additionally, we plot the 2$\mu$m medium-band colors of quiescent JAGUAR galaxies (black points) to demonstrate the capability of characterizing emission lines and the movement of the 4000\AA\ break into and out of the filters above $z>3$.

\begin{deluxetable*}{lllll}[!htbp]
\caption{ Estimated number of galaxies with robustly measured spectral features (Fig. \ref{fig:fluxlims}, left panel) in our NIRCam pointing.  
}\label{tab:counts}
\tablehead{\colhead{Filter}   & \colhead{Feature} &  \colhead{Redshift }& \colhead{Predicted N$_{obj}$$^{a}$ } & \colhead{Observed N$_{obj}$$^{b}$ }  \\[-0.3cm]
 \colhead {}    &  \colhead {}   &     \colhead {}   &  \colhead {}    \\[-0.4cm]}
 \startdata
F182M+F210M & H$\alpha$+\NII\ &   $1.5-2.5$ & 15$\pm$1 & 26  \\
& \OIII+H$\beta$ &   $2.4-3.5$  &  27$\pm$2 & 141 \\
\hline
F430M+F460M+F480M & H$\alpha$+\NII\ &    $5.3 - 6.6$ & 2$1\pm$1 & 67 \\
 &  $[\mathrm{OIII}]$+H$\beta$ &  $7.3-9.3$ &  4$\pm$1 & 16  \\
\enddata
\tablenotetext{a}{Predicted number of sources in our $\sim10$ square arcmin NIRCam coverage with JAGUAR above S/N $>$ 7 for emission line and continuum bands. 
For \Halpha+\NII\ and \OIII+Hb we show sources with medium-band colors $>$ 1 magnitude (roughly corresponding to EW $>$ 500\AA; see description in Section \ref{sec:empirical}).}
\tablenotetext{b}{Observed galaxies in our survey that meet the criteria listed in (a) for each set of emission lines.}
\end{deluxetable*}

In the 2nd and 4th panels we show for comparison the colors versus redshift for real galaxies detected in our medium-band imaging. To measure medium-band colors, we use the forced photometry for sources detected in the 5-band stack as outlined in Section \ref{sec:photometry}. 

We also measure photometric redshifts in our medium-band detected sample (Hainline et al. 2023, in preparation). To enable as accurate as possible redshifts, we make use of a suite of the deepest ancillary optical-near-infrared photometry available in this field, in addition to the new imaging from our medium-band survey (see Section \ref{sec:ancillary}). 
We use the redshift at the peak of the photometric redshift distribution measured using EAZY  \citep{Brammer2008}. We do not set priors and use the standard EAZY templates with a custom-supplemented set of SED templates that include  bluer continuums and stronger line and nebular continuum emission \citep[see also][]{Larson2022}. Where spectroscopic redshifts are available we replace the photometric estimate with the spectroscopic one.

We find that the simulated color distributions with redshift from JAGUAR are largely reproduced by real galaxies identified in our imaging. In particular, we find that  H$\alpha$+\NII\ and \OIII+\Hbeta\ drive the strongest colors at both 2$\mu$m and 4$\mu$m wavelengths. We also find examples where color excesses are measurable from both Paschen-$\alpha$ (see also Figure \ref{fig:SEDs}) and \SIII+\HeI, in particular at 4$\mu$m. Of those three lines, the prominent flux excesses in the yellow shaded region in the bottom panel of Figure \ref{fig:MBcolors} originate from the \HeI\ line. 

Taking these redshifts and colors at face value, we find that our survey likely identified 26 \Halpha+\NII\ emitters in our 2$\mu$m imaging at $z\sim$1.5-2.5 with EW$>$500\AA\ (corresponding roughly to absolute color difference $>$1 magnitudes). At z$\sim$5.5-6.6, we identified 67 likely \Halpha+\NII\ emitters with EW$>$500A in our 4$\mu$m imaging. These numbers are for detections with S/N in both filters of at least 7. 
Similarly, we identify 141 \OIII+\Hbeta\ emitters in our 2um imaging at $2.4<z<3.5$ and 16 in the 4$\mu$m imaging at $7.3 < z < 9.3$. 

 In Table \ref{tab:counts} we compare our observed number of \Halpha+\NII\ and \OIII+\Hbeta\ emitters across redshifts to those predicted by the JAGUAR phenomenological galaxy evolution model. For this comparison, we include JAGUAR sources whose fluxes are S/N $>$ 7 in both the emission line and continuum band as is done for the observed galaxies in Figure \ref{fig:MBcolors}. We find that JAGUAR generally underpredicts the number of galaxies with observed medium-band colors $>$1 magnitude for the redshift ranges where these emission lines enter our filters (see Table \ref{tab:counts}). This is in line with recent comparisons finding JAGUAR generally underpredicts strong \Halpha+\NII\ and \OIII+\Hbeta\  emitters, in particular up to $>8\times$ for   \OIII+\Hbeta\ at $z\sim8$ \citep{Maseda2019,DeBarros2019, Sun2022}. 
Possible explanations for the stronger observed lines include more stochastic star formation in real galaxies, poor representation of the tails of real galaxy property distributions that drive strong lines (e.g. LogU), and that more extreme stellar populations are not represented in JAGUAR.

\begin{figure*}[!ht]
\includegraphics[width=1\textwidth, trim=18 20 0 10, clip]{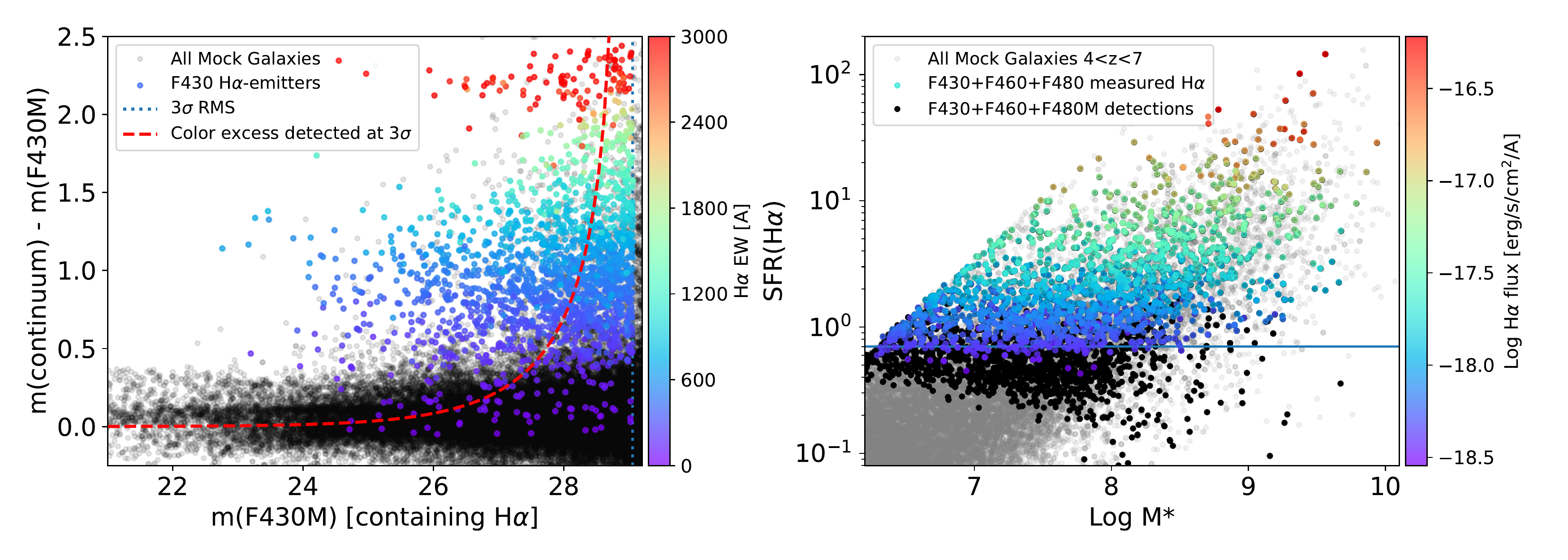}
\caption{Left panel: mock observations using JAGUAR simulating our 4$\mu$m-selected H$\alpha$ source population (colored points) from their color excess in our medium-bands. (example shown is for $5.3<z<5.7$ using H$\alpha$ emitters identified from F430M excess in one realization of JAGUAR; 120 sq arcmin). Continuum is measured using the other medium-bands (F460M, F480M). Rainbow points in excess of the red line have robustly measured color excesses \citep[$>3\sigma$; e.g.][]{Bunker1995}. Right panel: Region of the SFR-M* diagram where our survey can robustly measure H$\alpha$ from color excesses in all three 4$\mu$m filters (using criteria demonstrated in the left panel), which roughly corresponds to a limiting unobscured SFR traced by \Halpha\ of $\sim$0.7 M$_{\odot}$ yr$^{-1}$ where equivalent width is measurable (indicated by a horizontal line; number counts in Table \ref{tab:counts}).}\label{fig:fluxlims}
\end{figure*}

\subsection{Simulating emission line constraints using medium-band  imaging} \label{sec:simulations}

In this section, we translate our photometric limits into emission line equivalent width limits to understand the swath of the galaxy population for which we can constrain physical parameters. To make these translations, we make use of the NIRCam photometry and emission line properties of JAGUAR mock galaxies. In Figure \ref{fig:fluxlims},  we show the color excesses of JAGUAR galaxies where \Halpha+\NII\ enters the F430M band ($5.3<z<5.7$). To measure the continuum magnitude near the emission line, we use a linear fit with the F460M and F480M fluxes. The full mock galaxy distribution is shown in black points, while mock galaxies that would be detected at $>3\sigma$ in our imaging are shown as colors (color coded by their intrinsic equivalent width of H$\alpha$ emission). It is clear that as H$\alpha$ EW increases, so do the observed $4\mu$m color excesses. Galaxies with observed colors that are in excess of the red dashed line have \Halpha\ EWs that are measurable using colors at the $>3\sigma$ level by our survey \citep[e.g.][]{Bunker1995, Shioya2009, Sobral2013}. We find that we are sensitive to rest-frame \Halpha\ EW $\gtrsim$ 50\AA\ at brighter magnitudes ($<$24-26 AB), and near our detection limit, we can still identify sources with rest-frame EW $\gtrsim$ 2500\AA. 

In the right panel of Figure \ref{fig:fluxlims} we plot the SFR versus M$_{\star}$ for mock galaxies with EWs that can be measured as detectable color excesses at the $>3\sigma$ level (colored points). The right panel contains objects identified using the analogous selection for emission lines inside both F460M and F480M, in addition to F430M (which is presented as an example in the left plot). In this right panel, we convert the intrinsic \Halpha\ luminosity (as observed, uncorrected for dust attenuation) of the detectable mock galaxies to an unobscured SFR(\Halpha) using \citealt{Kennicutt1998}, converted to \citealt{Chabrier2003} initial mass function (IMF). We find that given our deep imaging limits,  the galaxies for which we are able to robustly measure color excesses from H$\alpha$ includes galaxies with SFR as low as $\gtrsim0.7$ M$_{\odot}$/year (right panel of Figure \ref{fig:fluxlims}).  

Given our imaging detection limits, this includes galaxies with line fluxes as low as LogF $> -18$ erg/s/cm$^2$ (but note that our sensitivity to line flux depend on apparent magnitude and is not a uniform limit). Compared to existing {\it JWST} grism observations (FRESCO with $5\sigma$ line flux sensitivity $\gtrsim2\times10^{-18}$ ergs/s/cm$^2$ between 4.3-4.6$\mu$m, Sun et al. personal communication), we find our imaging is roughly 2$\times$ deeper at 4$\mu$m in terms of emission line flux sensitivity (for EW with detectable colors).

This demonstrates our medium-band imaging can provide emission line and related physical parameter constraints for galaxies in new parameter space from across redshifts from $0.3<z\lesssim20$. 
Figure \ref{fig:fluxlims} shows the power of our dataset to constrain uncertain parameters such as SFR and emission line strength. This result is in line with analysis presented elsewhere demonstrating quantitatively that {\it JWST} medium-band imaging improves physical parameter recovery in high-redshift galaxies \citep[][]{RobertsBorsani2021}.

\section{Plans for release of higher-level data products}\label{sec:future}

Upon acceptance, we will make publicly available our first data release that includes version 1 of our science-ready mosaics, enabling the community to begin exploiting this exciting dataset. In the future, we plan additional data product releases, including science-ready photometric catalogs with value-added parameters. Additionally, these data will also be integrated into the JADES survey data products once they are publicly released (expected October 2023), including mosaics that are astrometrically aligned with and catalogs incorporating existing {\it HST} data.

\section{Summary}

We have planned and executed JEMS, a 5 filter medium-band survey with {\it JWST} in Cycle 1 at 2 and 4$\mu$m wavelength. Our survey includes F182M, F210M, F430M, F460M, and F480M images taken with NIRCam over the HUDF and coordinated parallel NIRISS images using F430M and F480M that fall on the CANDELS footprint in GOODS-S. In this work we have demonstrated that our data are capable of measuring and characterizing strong emission lines at all redshifts from $0.3<z\lesssim20$, opening the door to better constrained physical parameters in high-redshift galaxies. Medium-band imaging with {\it JWST} represents a new and exciting resource with high efficiency compared to spectroscopy, enabling new science across redshifts as demonstrated in this paper.

\appendix

To facilitate joint analysis of our new JWST imaging with existing {\it HST} imaging (e.g. HLF, \citealt{Whitaker2019}) we document here the astrometric adjustments that we performed to public mosaics available elsewhere. As described in Section \ref{sec:data}, the astrometry of our mosaics is tied to the GAIA system. Therefore one can co-analyze with the CHArGE imaging (although we note that their 100mas pixel scale is different from ours, which are 30mas pixels).

Alternatively, one can use the HLF imaging (v2.0) with the same 30mas pixel scale (and the same nominal pixel registration) if the HLF image header information is updated. We provide these updates here, which involve both a reference pixel offset and a slight rescaling of pixel size. 
To adjust the pixel size, we update the {\tt CRVAL1} and {\tt CRVAL2} values by $-4.08$ and $+2.76$ mas respectively. To rescale the pixel size, we scale the {\tt CD1\_1} and {\tt CD2\_2} values by 0.9998258, 0.9998248 respectively. We will provide a python script on our data release page (to be released upon acceptance) that will perform these adjustments to the HLF images and catalog coordinates.

\begin{acknowledgments}

We thank Gabe Brammer for making his reduction of our data available to us and to the community, and Ivo Labbe, Kate Whitaker for useful discussions. We also thank Armin Rest, Mario Gennaro, Jarron Leisenring, Everett Schlawin, and Bryan Hilbert for advice related to NIRCam medium-band image processing. 
This work is based in part on observations made with the NASA/ESA/CSA James Webb Space Telescope. The data were obtained from the Mikulski Archive for Space Telescopes at the Space Telescope Science Institute, which is operated by the Association of Universities for Research in Astronomy, Inc., under NASA contract NAS 5-03127 for JWST. These observations are associated with JWST
Cycle 1 GO program \#1963. Support for program
JWST-GO-1963 was provided by NASA through a grant
from the Space Telescope Science Institute, which is operated by the Associations of Universities for Research in
Astronomy, Incorporated, under NASA contract NAS 5-
26555. The authors acknowledge the FRESCO team led by PI Pascal Oesch for developing their observing program with a zero-exclusive-access period. The work of CCW is supported by NOIRLab, which is managed by the Association of Universities for Research in Astronomy (AURA) under a cooperative agreement with the National Science Foundation. S.A. acknowledges support from the James Webb Space Telescope (JWST) Mid-Infrared
Instrument (MIRI) Science Team Lead, grant 80NSSC18K0555, from NASA Goddard Space Flight Center to the University of
Arizona. 

\end{acknowledgments}

\software{astropy \citep{Astropy2013,Astropy2022},  
          Cloudy \citep{Ferland2013}, 
          photutils \citep{photutils}
          WebbPSF \citep{Perrin2015ascl.soft04007P}
          }

\bibliography{manu}{}
\bibliographystyle{aasjournal}

\end{document}